\documentclass[trackchanges,twocolumn,twocolappendix]{aastex701}

\newcommand{\jh}[1]{{\color{black}#1}}

\shorttitle{H$\alpha$ of SR 12 c}
\shortauthors{Hashimoto et al.}

\graphicspath{{./}{figures/}}

\begin{document}

\title{Asymmetric, variable H$\alpha$ line profile in planetary mass object SR 12 c}


\author[0000-0002-3053-3575]{Jun Hashimoto}
\affil{Academia Sinica Institute of Astronomy \& Astrophysics (ASIAA), 11F of Astronomy-Mathematics Building, AS/NTU, No.1, Sec. 4, Roosevelt Rd., Taipei 106319, Taiwan}
\affil{Astrobiology Center, National Institutes of Natural Sciences, 2-21-1 Osawa, Mitaka, Tokyo 181-8588, Japan}
\email[show]{jhashimoto@asiaa.sinica.edu.tw}  

\author[0000-0003-0568-9225]{Yuhiko Aoyama}
\affil{School of Physics and Astronomy, Sun Yat-sen University, Guangdong 519082, People’s Republic of China}
\email{}

\author[0000-0001-9248-7546]{Michihiro Takami}
\affil{Academia Sinica Institute of Astronomy \& Astrophysics (ASIAA), 11F of Astronomy-Mathematics Building, AS/NTU, No.1, Sec. 4, Roosevelt Rd., Taipei 106319, Taiwan}
\email{}

\author[0000-0003-3882-3945]{Shinsuke Takasao}
\affil{Department of Earth and Space Science, Graduate School of Science, The University of Osaka, Toyonaka, Osaka 560-0043, Japan}
\affil{Humanities and Sciences/Museum Careers, Musashino Art University, Tokyo 187-8505, Japan}
\email{}

\begin{abstract}
Young, forming planetary-mass objects often exhibit clear signatures of ongoing mass accretion and are thought to accrete material through processes analogous to those operating in young stars. In this study, we present high-spectral-resolution observations of asymmetric and time-variable H$\alpha$ line profiles from the planetary-mass companion SR~12~c. The H$\alpha$ line was observed at a resolving power of $R \sim 49,000$--40{,}000 (corresponding to 6.1--7.5~km~s$^{-1}$) using the High Dispersion Spectrograph (HDS) on the 8.2 m Subaru Telescope. Strong H$\alpha$ emission is clearly detected, while higher-order Balmer lines (H$\beta$, H$\gamma$, and H$\delta$) are not detected due to their faintness. The H$\alpha$ line profiles are well spectrally resolved and exhibit blueshifted emission peaks, which can be interpreted as arising from either (a) emission partially absorbed by redshifted accreting material along the line of sight and/or (b) geometric occultation by the inner circumplanetary disk. Moreover, the H$\alpha$ flux shows significant variability at 43.6~$\pm$~6.4~\% relative to the peak flux on hourly timescales. During a continuous 2.5-hour observing sequence, the emission component peaking at approximately $-30$~km~s$^{-1}$ weakened over the first hour. Subsequently, an emission component centered near $-10$~km~s$^{-1}$ became dominant and remained stable for the remaining 1.5 hours. We discuss possible interpretations of this behavior. Overall, these results support that magnetospheric accretion is operating in the planetary-mass object SR~12~c \jh{while a scenario combining boundary-layer accretion with a failed wind cannot be ruled out.}

\end{abstract}


\section{Introduction} \label{sec:intro}

In the late stages of planet formation, when the accreting planetary-mass objects become visible at optical and near-infrared wavelengths due to the dispersal of their surrounding material, mass accretion signatures---such as UV continuum excess and line emission---have been observed \citep[e.g.,][]{Joergens2013OTS44,Zhou14HST,Haffert2019PDS70bc,Zhou2021PDS70,Zhou2022ABAurb,Zhou2023ABAurb,Zhou2025PDS70c,Currie2025ABAurb,Almendros-Abad2025Cha1107-7626}. These accretion tracers have been widely used to investigate whether accretion processes in planetary-mass objects are analogous to those operating in young stars \citep[e.g.,][]{Aoyama2019,Hashimoto2020MUSE,Viswanath2024-2m1115,Demars2026D1ABb}.

\begin{deluxetable*}{ccccccc}[t!]
\tablewidth{0pt}
\tablecaption{HDS Observing log of SR 12 c\label{tab:obs}}
\tablehead{
\colhead{Observing sequence} & \colhead{Time UTC$^\dagger$} & \colhead{Object} & \colhead{Exposure time} & \colhead{Airmass} & \colhead{Spatial FWHM$^\ddagger$} & \colhead{Spectral resolution$^*$} \\
\colhead{}          & \colhead{}     & \colhead{}       & \colhead{(second)} & \colhead{}        & \colhead{($\arcsec$)} 
}
\startdata
\#1 & 6:23 & SR 12 c & 1,800 & 1.46 & 0.73 & 49,315\\
\#2 & 6:54 & SR 12 c & 1,800 & 1.55 & 0.84 & 42,857\\
\#3 & 7:25 & SR 12 c & 1,800 & 1.68 & 0.88 & 40,909\\
\#4 & 7:55 & SR 12 c & 1,800 & 1.88 & 0.86 & 41,860\\
\#5 & 8:26 & SR 12 c & 1,800 & 2.20 & 0.90 & 40,000\\
\#6 & 8:58 & Sky     & 596   & 2.20 & \nodata \\
\enddata
\tablecomments{
$\dagger$Observations were carried out on 2025 August 9 (UTC). 
$\ddagger$The spatial FWHM of SR~12~c was measured along the spatial axis of the two-dimensional spectrum around the H$\alpha$ wavelength. 
$*$Since the spatial resolution was finer than the 1\arcsec\ slit width, the effective spectral resolution was set by the spatial resolution.
} 
\end{deluxetable*}

Magnetospheric accretion is a leading mechanism for mass accretion in young stars \citep[e.g.,][]{Hartmann2016accretion}. In T Tauri stars, broad emission lines and/or redshifted absorption components---known as inverse P-Cygni profiles---have been observed in various systems \citep[e.g.,][]{Bertout1982Accretion,Edwards1994Accretion,Muzerolle1998Accretion} and are generally interpreted as signatures of accreting inflows \citep[e.g.,][]{Calvet1992Model,Hartmann1994Model,Muzerolle1998Model,Wilson2022Model}. Similar features have also been reported in accreting planetary-mass objects \citep[e.g.,][]{Demars2023Variability,Currie2025ABAurb}. However, the formation of redshifted absorption components depends sensitively on radiative transfer and geometric effects \citep{Calvet1992Model,Hartmann1994Model,Kurosawa2006Halpha}, implying that such features may not always be observed \citep{Edwards1994Accretion}.

In addition to redshifted absorption components, asymmetric line profiles---so-called blueward asymmetry \citep[e.g.,][]{Kurosawa2006Halpha}---can also arise from geometric occultation by the inner disk. This is because the fraction of red-shifted accretion flow obscured by the inner disk is larger than that of blue-shifted accretion flow obscured by the photosphere. Redshifted absorption and blueward asymmetry are often confused, but they are physically distinct phenomena. Overall, both effects are natural consequences of magnetospheric accretion.

Emission-line variability is widely observed in young accreting stars, spanning timescales from sub-hours to years \citep[e.g.,][]{Fischer2023PPVII_variability,Hartmann2016accretion,Bouvier2007AATau,Alencar2012variability}. While some variability patterns reflect stellar rotation or the influence of intervening dust, variations in intrinsic emission-line strength are driven by unstable accretion driven by magnetospheric instabilities, and clumpy or time-variable accretion inflows from the disk onto the star. Variability has also been reported in accreting planetary-mass objects \citep[e.g.,][]{Demars2023Variability,Demars2026D1ABb,Zhou2025PDS70c,Close2025WISPIT2b-Halpha,Almendros-Abad2025Cha1107-7626}. Although the origin of accretion variability in such objects remains unclear, changes in circumplanetary disk properties or transient accretion bursts may be responsible for the observed behavior.

In this paper, we report another example of a variable, accreting planetary-mass object, SR~12~c, designated in SIMBAD as NAME~SR~12~AB~c, (spectral type: L0; $T_{\rm eff} = 2600 \pm 100$~K; \citealp{Santamaria-Miranda2018SR12c}; mass: $11 \pm 3~M_{\rm Jup}$; radius: $1.6~R_{\rm Jup}$; \citealp{Wu2022SR12c}; age: $\sim$2 Myr; \citealp{Kuzuhara2011SR12c}; distance: 112.3 pc; \citealp{Gaia2016,Gaia2023}; \jh{systemic velocity\footnote{\citet{Santamaria-Miranda2018SR12c} measured the systemic velocity of SR~12~c to be $-6.1 \pm 1.0$~km~s$^{-1}$ from the calcium triplet emission lines and $-7.4 \pm 2.8$~km~s$^{-1}$ from the sodium absorption doublet. Taking the arithmetic mean of these measurements and propagating their uncertainties yields a systemic velocity of $-6.8 \pm 3.0$~km~s$^{-1}$.}: $-6.8 \pm 3.0$~km~s$^{-1}$}; \citealp{Santamaria-Miranda2018SR12c}), located in the $\rho$ Ophiuchi star-forming region.

Previous observations suggest that SR~12~c is surrounded by a dusty, accretion disk. Thermal dust emission has been detected at both mid-infrared \citep{Martinez2022Spitzer-CPD} and sub-millimeter wavelengths \citep{Wu2022SR12c}. The dust disk mass is estimated to be 0.012--0.054 $M_{\oplus}$ \citep{Wu2022SR12c}, depending on the assumed opacity. Multiple hydrogen lines have been detected, indicating that SR~12~c is accreting at a rate of $10^{-11.08\pm0.40}$~$M_{\sun}$~yr$^{-1}$ \citep{Santamaria-Miranda2018SR12c}. A Balmer continuum excess at $\lambda \lesssim 364.6$~nm has also been detected, providing an independent estimate of the accretion rate of $10^{-11.10\pm0.10}$~$M_{\sun}$~yr$^{-1}$ \citep{Finley2026arXiv}.

SR~12~c is separated from the central binary SR~12~AB by 8\farcs7 (980~au), making it one of the widest-separation planetary-mass companions known. For closer companions at sub-arcsecond separations \citep[e.g., the PDS 70 planets;][]{Keppler2018PDS70b}, high-contrast instruments are typically required to suppress contamination from the central stars. In contrast, given the ultrawide separation of SR~12~c, such instruments are unnecessary, making SR~12~c an excellent target for characterizing accretion signatures with high-resolution spectroscopy. 

\section{Observations} \label{sec:obs}

We carried out high-spectral-resolution observations with Subaru/HDS \citep{Noguchi2002HDS} on 2025 August 9 (UT), under program ID S25B-064 (PI: Y.~Aoyama). SR~12~c was observed in five sequential 30-minute exposures to study the Balmer lines. A sky frame was also obtained for 596~s after the observations of SR~12~c. The sky was clear, and the spatial resolution (FWHM) of SR~12~c was approximately 0\farcs7–0\farcs9. The observations used the SETUP-Yd configuration, covering wavelengths from 403 to 674~nm. As shown in Table~\ref{tab:obs}, because the spatial resolution was finer than the 1\arcsec\ slit width, the effective spectral resolution was $R \sim 49,000$--40{,}000 (6.1--7.5~km~s$^{-1}$). Our observing setup covers the Balmer lines H$\alpha$ (656.3~nm in air), H$\beta$ (486.1~nm), H$\gamma$ (434.1~nm), and H$\delta$ (410.2~nm). The observing log is summarized in Table~\ref{tab:obs}.

Given the faintness of SR~12~c ($R \sim 20.8$ mag), direct tracking was not feasible. Instead, we used the host binary SR~12~AB as the guide star. We then iteratively adjusted the position of SR~12~c on the slit, using its H$\alpha$ flux to identify the optimal placement where the emission is maximized.

The data were reduced using the HDS reduction procedure \verb#hdsql#, in conjunction with the Image Reduction and Analysis Facility (IRAF; \citealt{Tody1986IRAF}), as provided by the observatory\footnote{\url{https://www.naoj.org/Observing/Instruments/HDS/hdsql-e.html}}. The reduction process included overscan subtraction, scattered-light and cross-talk correction, nonlinearity correction, flat-fielding, and cosmic-ray removal.

\begin{figure}[htbp]
    \includegraphics[width=\linewidth]{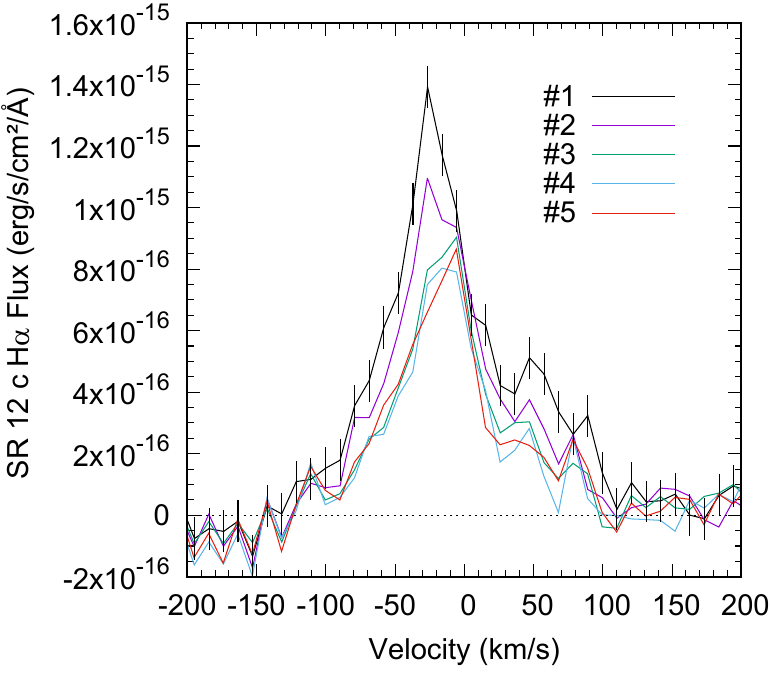} 
    \caption{
    H$\alpha$ spectra of SR~12~c from five individual datasets. All spectra show blueshifted peaks and are asymmetric with respect to \jh{the systemic velocity of $-6.8 \pm 3.0$~km~s$^{-1}$} (\citealp{Santamaria-Miranda2018SR12c}; see also Figure~\ref{fig:sr12c-each-spec}). While spectra \#1 and \#2 show peaks at approximately $-30$~km~s$^{-1}$, the subsequent spectra show peaks at approximately $-10$~km~s$^{-1}$. A difference of $-20$~km~s$^{-1}$ is significant. While the velocity resolution in our observations are 6.1--7.5~km~s$^{-1}$, the data were binned to a velocity resolution of 10.5~km~s$^{-1}$ to increase SNR. Unidentified systematic effects may be present at approximately $-175$ and $+175$~km~s$^{-1}$ (see \S~\ref{sec:obs}). The error bars are shown only for the first spectrum for presentation purposes. The absolute flux is provided as a reference (see \S~\ref{sec:obs}).
    }\label{fig:sr12c-spectrum}
\end{figure}

Spectral extraction was performed using the \verb#apall# task in IRAF. The sky background was subtracted from each extracted spectrum. As the photometric standard star HR~7001 was saturated in our observations, we instead used the telluric standard star HD~227634 to derive the absolute flux calibration. The visible spectrum of HD~227634 provided by Gaia \citep{Gaia2016,Gaia2023} was used for this purpose. The resulting absolute flux values are provided as a reference, given the presence of unknown systematics. Blaze-curvature correction was applied using nearby spectral orders from the HD~227634 data, as the Balmer absorption features in this star are strong. Wavelength calibration was performed using a Th–Ar comparison lamp frame. Finally, the spectra were corrected for the heliocentric radial velocity. While the RMS of the wavelength calibration derived from the Th–Ar comparison was $3.4\times10^{-3}$~\AA, corresponding to 0.16~km~s$^{-1}$ at H$\alpha$, the accuracy of the RV correction performed with IRAF is reported to be 0.005~km~s$^{-1}$ in the official documentation\footnote{\url https://iraf.readthedocs.io/en/doc-autoupdate/index.html}.
\jh{In addition, we confirmed that the wavelengths of the sky lines detected in our data agree with those in the sky-line list provided by \citet{Hanuschik2003Skylines} within 0.5~km~s$^{-1}$.}

Figure~\ref{fig:sr12c-spectrum} shows the H$\alpha$ spectra of SR~12~c. The residuals at approximately $-175$ and $+175$~km~s$^{-1}$ show weak negative and positive features, respectively. However, we confirmed that the overall residuals are consistent with zero over a wider velocity range of $-1,000$ to $+1,000$ km s$^{-1}$ (not shown here); in other words, there is no systematic trend in the residuals. Nevertheless, unidentified systematic effects may still be present at approximately $-175$ and $+175$ km s$^{-1}$. Similar weak systematic features are also visible, for example, at approximately $-900$, $-550$, and $+900$ km s$^{-1}$. We do not correct these systematic effects using polynomial fitting because the correction is not expected to significantly alter the results presented below.

\section{Results} \label{sec:results}

While our observing setup covers multiple hydrogen lines, only H$\alpha$ was detected, even after stacking all five datasets. Additionally, no continuum emission was detected. Figure~\ref{fig:sr12c-spectrum} shows the H$\alpha$ spectra of SR~12~c from the five individual datasets. All spectra exhibit blueshifted peaks \jh{relative to the systemic velocity of SR~12~c ($-6.8 \pm 3.0$~km~s$^{-1}$; \citealp{Santamaria-Miranda2018SR12c})}. Spectra \#1 and \#2 show peaks at approximately $-30$~km~s$^{-1}$, whereas the subsequent spectra show peaks at approximately $-10$~km~s$^{-1}$. Since the velocity resolution in our observations are 6.1--7.5~km~s$^{-1}$, a difference of $-20$~km~s$^{-1}$ is significant. The spectral wings at approximately $-100$~km~s$^{-1} < v < -50$~km~s$^{-1}$ and $+50$~km~s$^{-1} < v < +100$~km~s$^{-1}$ decrease from \#1 to \#3 and then remain stable thereafter. The H$\alpha$ fluxes listed in Table~\ref{tab:flux} follow this trend within the uncertainties. Specifically, the H$\alpha$ flux decreases by 43.6~$\pm$~6.4~\% from \#1 to \#4 over a timescale of $\sim$1 hour, while the variability between \#3 and \#5 is negligible. The 43.6~$\pm$~6.4~\% flux amplitude relative to the peak flux could represent a lower limit. Additional observations over longer timescales are required to investigate larger-amplitude variability, \jh{as our observations may not have spanned a full range of variability}.

\begin{figure*}[htbp]
    \includegraphics[width=\linewidth]{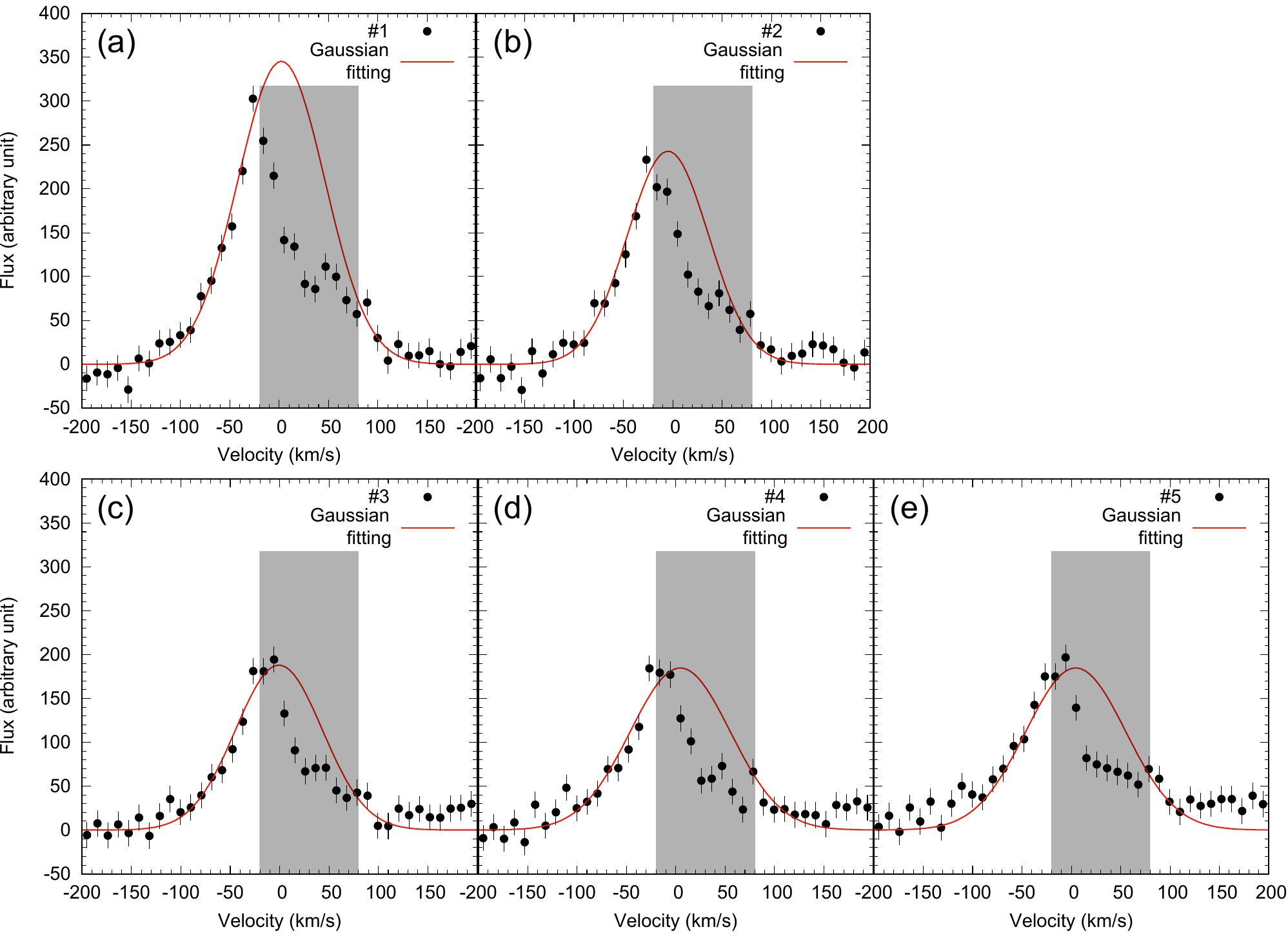} 
    \caption{
    Same as Figure~\ref{fig:sr12c-spectrum}, but showing the individual H$\alpha$ spectra of SR~12~c. Panels (a) to (e) correspond to observing sequences \#1 to \#5, respectively. The red lines represent the results of Gaussian fitting to the spectral wings at $-$200~km~s$^{-1}$~$<v<$~$-$20~km~s$^{-1}$ and $+$80~km~s$^{-1}$~$<v<$~$+$200~km~s$^{-1}$. The gray regions at $-$20~km~s$^{-1}$~$<v<$~$+$80~km~s$^{-1}$ were excluded from the Gaussian fitting. The fitting was performed using {\tt gnuplot} (\url{http://www.gnuplot.info}). Unidentified systematic effects may be present at approximately $-175$ and $+175$ km s$^{-1}$ (see \S~\ref{sec:obs}).
    }\label{fig:sr12c-each-spec}
\end{figure*}

In Figure~\ref{fig:sr12c-each-spec}, we show these spectra in different panels. All spectra are asymmetric with respect to \jh{the systemic velocity at approximately $-7$~km~s$^{-1}$}. Assuming that the asymmetry is caused by redshifted absorption, we performed Gaussian fitting to the spectral wings at $-$200~km~s$^{-1}$~$<v<$~$-$20~km~s$^{-1}$ and $+$80~km~s$^{-1}$~$<v<$~$+$200~km~s$^{-1}$. The resulting FWHM and peak velocity are summarized in Table~\ref{tabA:gaussian-fitting}. The fitted peak velocities are approximately 0~km~s$^{-1}$, while the Gaussian FWHMs are approximately 100~km~s$^{-1}$. In Gaussian fitting, we do not include a continuum-level offset term. Therefore, although the systematic effects mentioned in \S~\ref{sec:obs} are not corrected, the correction is not expected to significantly alter the results presented here.

For accreting stars, the line width is typically comparable to the free-fall velocity \citep[e.g.,][]{Hartmann2016accretion}. Line widths exceeding the free-fall velocity can result from Stark broadening \citep[e.g.,][]{Muzerolle2001Model}, whereas narrower line widths may arise when the magnetospheric truncation radius is relatively small. The line width of approximately 100~km~s$^{-1}$ for SR~12~c is smaller than its free-fall velocity from infinity, which is approximately 130--180~km~s$^{-1}$. A similar trend has also been reported for the planetary-mass companion TWA~27~B \citep{Marleau2024TWA27B} and recent studies of T~Tauri stars \citep{Pittman2025_ODYSSEUS-1}. Using a smaller truncation radius of, for example, 3 $R_{*}$ (4.8 $R_{\rm Jup}$ for~SR 12~c), as is found in T~Tauri stars  \citep{Pittman2025_ODYSSEUS-1}, the lower limit on the free-fall velocity of SR~12~c becomes approximately 110~km~s$^{-1}$. Therefore, the measured line width in Table~\ref{tabA:gaussian-fitting} may be consistent with the free-fall velocity expected from magnetospheric accretion in SR 12 c.

\begin{deluxetable*}{cccc}[hbpt!]
\tablewidth{0pt}
\tablecaption{H$\alpha$ flux of SR 12 c\label{tab:flux}}
\tablehead{
\colhead{Observing sequence} & \colhead{Integrated H$\alpha$ flux} & \colhead{Corrected H$\alpha$ flux}  & \colhead{Throughput}\\
\colhead{}  & \colhead{(10$^{-15}$ erg s$^{-1}$ cm$^{-2}$)} & \colhead{(10$^{-15}$ erg s$^{-1}$ cm$^{-2}$)} & \colhead{(\%)} \\
\colhead{(1)} & \colhead{(2)} & \colhead{(3)}  & \colhead{(4)}
}
\startdata
\#1 & 2.65 $\pm$ 0.10 & 2.98 $\pm$ 0.11 & 89.0\\
\#2 & 2.00 $\pm$ 0.10 & 2.40 $\pm$ 0.12 & 83.3\\
\#3 & 1.58 $\pm$ 0.12 & 1.94 $\pm$ 0.15 & 81.3\\
\#4 & 1.38 $\pm$ 0.15 & 1.68 $\pm$ 0.18 & 82.3\\
\#5 & 1.44 $\pm$ 0.17 & 1.79 $\pm$ 0.21 & 80.3\\
\enddata
\tablecomments{(1) Observing sequence usd in Table~\ref{tab:obs}. (2) The absolute flux values are provided as a reference (see \S~\ref{sec:obs}), while the relative flux values are robust. (3) The throughput (i.e., slit loss), listed in column (4), has been taken into account. (4) The throughput represents the fraction of light that passes through the 1\arcsec\ slit under the seeing conditions listed in Table~\ref{tab:obs}.} 
\end{deluxetable*}

\begin{deluxetable}{ccc}
\tablewidth{0pt}
\tablecaption{FWHM and peak velocity from the Gaussian fitting in Figure~\ref{fig:sr12c-each-spec}\label{tabA:gaussian-fitting}}
\tablehead{
\colhead{Observing sequence}  & \colhead{Peak velocity} & \colhead{FWHM}\\
\colhead{}        & \colhead{(km/s)}        & \colhead{(km/s)} 
}
\startdata
\#1 &    2.3 $\pm$ 2.1 & 103.8 $\pm$ 3.5\\
\#2 & $-$4.7 $\pm$ 2.3 &  96.7 $\pm$ 4.4\\
\#3 & $-$0.7 $\pm$ 3.5 & 103.8 $\pm$ 6.5\\
\#4 &    4.5 $\pm$ 3.9 & 119.5 $\pm$ 8.1\\
\#5 &    4.5 $\pm$ 3.9 & 119.5 $\pm$ 8.1\\
\enddata
\end{deluxetable}

\section{Discussions} \label{sec:discussions}

\subsection{Accretion process} \label{subsec:accretion}
For accreting stars, two accretion processes have been studied to better understand stellar accretion \citep[e.g.,][]{Bertout1988,Koenigl1991}. The first is boundary layer accretion, where the faster-rotating Keplerian disk connects with the slowly rotating star. Disk materials accrete onto the stellar surface through an equatorial shear boundary layer. The second is magnetospheric accretion, where strong magnetic fields truncate the protoplanetary disk and cause accretion flow from the disk inner edge to the stellar surface along the magnetic field lines.

Observations have shown that the permitted emission lines of T Tauri stars (e.g., H$\alpha$) are variable and asymmetric \citep[e.g.,][]{Edwards1994Accretion}. These line profiles cannot be explained by boundary layer accretion but instead support the magnetospheric accretion model \citep[][]{Hartmann1994Model}. Unless the stellar spin and magnetic axes are perfectly aligned, the accretion geometry changes with the stellar rotation \citep[e.g.,][]{Tessore2023}. Since protoplanetary disk gas accretes onto the hemisphere where the magnetic field axis is closer \citep{Romanova2003}, the accretion flow rarely interrupts the line of sight to the stellar surface when the magnetic axis is located behind the rotation axis. In contrast, when the magnetic axis is in front of the rotation axis, the accretion flow overlaps with the stellar surface, resulting in a spectral profile with a strong redshifted absorption. In addition to such redshifted absorption, blueward asymmetry can also arise from geometric occultation by the inner disk in the context of magnetospheric accretion \citep[e.g.,][]{Kurosawa2006Halpha}. SR~12~c's H$\alpha$ spectra could also exhibit redshifted absorption or blueward asymmetry (Figure~\ref{fig:sr12c-spectrum} and \ref{fig:sr12c-each-spec}), which is consistent with the magnetospheric accretion model.

When a protoplanet is embedded within a protoplanetary disk, photometric variability is also expected in the case of boundary-layer accretion \citep{Takasao2021Halpha}. This arises because the surface accretion layers in the circumplanetary disk can be highly variable, leading to time-dependent accretion rates and corresponding variations in the H$\alpha$ luminosity of the protoplanet. However, since SR~12~c is not embedded within a protoplanetary disk, this scenario does not apply.

Meanwhile, boundary-layer accretion may also produce asymmetric line profiles. Although redshifted absorption by accretion flows is not expected in this scenario, a disk wind that fails to escape and returns to the inner disk---so-called ``failed wind'' \citep{Takasao2022MHD3DMA}---could generate asymmetric line profiles when it moves away along the line of sight. As discussed below, the expected emission-line width from boundary-layer accretion and the expected velocity of redshifted absorption caused by failed disk winds are of order $\sim$100~km~s$^{-1}$ and less than approximately 60-80~km~s$^{-1}$, respectively. These values are in good agreement with the line profile observed in SR~12~c (Figures~\ref{fig:sr12c-spectrum} and \ref{fig:sr12c-each-spec}). 

The Keplerian velocity at the inner disk boundary of SR~12~c is estimated to be approximately 100–130~km~s$^{-1}$. Therefore, emission lines originating from the boundary layer may exhibit line widths of order $\sim$100~km~s$^{-1}$. In the context of boundary-layer accretion $+$ failed wind, we assume that the planet has a weak magnetic field, leading to the formation of an incomplete magnetosphere. Assuming a magnetospheric radius of five times the radius of SR~12~c (5~$R_{*}$; 8.0~$R_{\rm Jup}$), comparable to the typical magnetospheric truncation radius of T~Tauri stars \citep[e.g.,][]{Hartmann2016accretion}, the velocity of the failed wind launched from the inner disk at $r\sim8.0$~$R_{\rm Jup}$ and returning to the same location should be lower than its escape velocity, which is approximately 60-80~km~s$^{-1}$. 

If the absorption by the failed wind occurs near the inner disk connected to the magnetosphere, the Keplerian timescale at $r \sim 8.0$~$R_{\rm Jup}$ is approximately 1~day. For absorption occurring beyond the magnetosphere, the corresponding variability timescale would be expected to be on the order of a few days. This is likely too long to account for the observed hourly variability of SR~12~c. \jh{However, three-dimensional MHD simulations suggest that the failed wind is turbulent and time-variable \citep{Takasao2022MHD3DMA}, which could introduce stochastic variations in the observed absorption. Therefore, a scenario combining boundary-layer accretion with a failed wind cannot be ruled out.}

\subsection{Variability in line emission} \label{subsec:variability}

Accretion variability mechanisms, including episodic, periodic, and stochastic variability, have been summarized in a recent review by \citet{Fischer2023PPVII_variability}. Since the 2.5-hour duration of our observations may not capture the full range of variability, it is difficult to distinguish between these variability mechanisms. Here, we discuss several possible interpretations.

One simple interpretation is that the variability is related to rotational modulation of the planet. When viewed at a nearly edge-on inclination, line-emitting surfaces of accreting objects (e.g., accretion shock regions) may be periodically occulted by the planet itself as it rotates. In addition, geometric projection effects can cause the line centers to shift over the rotation cycle. Although the exact spin period of SR~12~c is unknown, other young gas giant exoplanets, such as TWA~27~b \citep[$\sim$11 hours;][]{Zhou2016TWA27b} and $\beta$~Pic~b \citep[$\sim$8 hours;][]{Snellen2014betaPicb}, exhibit rotation periods of $\sim$10~hours. Assuming a similar rotation period for SR~12~c, our 2.5-hour observations would cover roughly one quarter of a full rotational cycle. Meanwhile, when the rotation velocity is regulated by the inner disk connected to the magnetosphere with 5~$R_*$ \citep[e.g.,][]{Hartmann2016accretion}, the rotation period of SR~12~c is $\sim$1~day (see \S~\ref{subsec:accretion}). 

If rotation were responsible for the observed variability, accreting funnel flows near the outer rim regions of SR~12~c would rotate with SR~12~c and eventually accrete onto its central regions. In this scenario, the emission peaks could shift from $\sim -30$~km~s$^{-1}$ to $\sim -10$~km~s$^{-1}$. However, rotational modulation would be expected to produce smooth, continuous changes in the velocity peaks, which are not observed. We therefore disfavor the rotational modulation scenario.

Meanwhile, short-timescale accretion variability is thought to originate from spatially and temporally inhomogeneous accreting inflows. Rather than occurring through a single, steady accretion column, material is channeled along multiple accretion columns that differ in density, geometry, and temporal behavior. As a result, individual accretion streams impact the star at different times and positions, producing stochastic and time-dependent variations in the observed accretion signatures \citep[e.g.,][]{Ingleby2013Multiple-flows,Takasao2022MHD3DMA,Ji2026TWHya}.

Another possibility is that SR~12~c is a tight binary planet. In this scenario, SR~12~c may consist of two accreting planets, each contributing to one of the observed H$\alpha$ line peaks. One of the components associated with the $\sim -30$~km~s$^{-1}$ peak may have faded during our observations, leaving the component at $\sim -10$~km~s$^{-1}$ to dominate the H$\alpha$ emission. To test this hypothesis, high-angular-resolution observations are required. Previous observations of SR~12~c with spatial resolutions poorer than $0\farcs1$ did not reveal any evidence for binarity \citep{Kuzuhara2011SR12c,Wu2022SR12c}. Further high-resolution observations are therefore necessary to test the binary-planet scenario.

Variability may also be induced by a change in the dominant emission mechanism. Two models have been developed to interpret H$\alpha$ emission from accreting planets: emission from the accretion flow \citep{Kwan2011Flow,Thanathibodee2019PDS70b}, which is thought to be the primary origin of H$\alpha$ emission lines in accreting stars \citep{Hartmann2016accretion}, and emission from the shock region \citep{Aoyama2018}. 
Such a transition in the dominant emission mechanism has been reported in accreting substellar objects \citep{Hashimoto2025accreton}. Since line ratios are key diagnostics for distinguishing between these models, simultaneous detection of multiple hydrogen lines is required. Based on multiple hydrogen emission lines of SR~12~c obtained simultaneously in 2016 \citep{Santamaria-Miranda2018SR12c}, Aoyama et al. (submitted) found that the dominant emission mechanism is shock-dominated. Future monitoring observations with VLT/XSHOOTER will be well suited to explore this scenario.

\subsection{Comparison with other sources} \label{subsec:comparison}

The number of high-resolution spectroscopic observations of accreting planetary-mass objects is increasing. Assuming typical line widths of 50--150~km~s$^{-1}$, roughly comparable to the free-fall velocity of 1–10 $M_{\rm Jup}$ planets, high-spectral-resolution spectroscopy with $R \gtrsim 10{,}000$ (corresponding to $\sim$30~km~s$^{-1}$) is requested at a minimum for resolving line profiles. In particular, this study reveals distinct velocity peaks at $\sim -30$ km s$^{-1}$ and $\sim -10$ km s$^{-1}$. Resolving such features requires a spectral resolution of at least $R \gtrsim 30{,}000$.

To the best of our knowledge, H$\alpha$ emissions in SR~12~c \citep{Santamaria-Miranda2018SR12c}, 2MASS~J11151597$+$1937266 (2M1115; \citealt{Viswanath2024-2m1115}), Cha~J11070768$-$7626326 (J1107; \citealt{Almendros-Abad2025Cha1107-7626}), and Delorme~1~(AB)b \citep{Demars2026D1ABb} have been observed with high-spectral-resolution spectroscopy. While J1107 shows clear redshifted absorption, 2M1115 and Delorme~1~(AB)b may also exhibit signatures of redshifted absorption in their asymmetric, tilted plateau profiles. Additionally, AB~Aur~b also shows redshifted absorption in its H$\alpha$ profile \citep{Currie2025ABAurb}. Although the current sample of accreting planetary-mass objects is still limited, these results suggest that the accretion process in planetary-mass objects is similar to that in stellar accretion, i.e., magnetospheric accretion.

SR~12~c shows an hourly variability at the $\gtrsim$45~\% level (\S~\ref{sec:results}; Table~\ref{tab:flux}). However, the number of reported cases of hourly variability in planetary-mass objects remains limited. Compared to SR~12~c, Delorme~1~(AB)b shows a smaller H$\alpha$ variability amplitude of $\sim$8~\% \citep{Demars2026D1ABb}. In addition, GQ~Lup~b and GSC~06214$-$00210~b exhibit variability amplitudes of $\sim7\pm7$~\% and $\sim20\pm17$~\%, respectively \citep{Demars2023Variability}, although these measurements are based on Pa$\beta$ emission. Among these objects, SR~12~c shows a significantly larger H$\alpha$ variability amplitude. Further observations are required to investigate whether these objects share a common emission mechanism.

\subsection{Comparison with \citet{Santamaria-Miranda2018SR12c}} \label{subsec:sr12c}

We briefly compare our results with previous observations of SR~12~c by \citet{Santamaria-Miranda2018SR12c}. They observed H$\alpha$ emission from SR~12~c with VLT/X-shooter \citep{Vernet2011Xshooter} on May 2, 2016. Although they used a slit width of 1\farcs5, corresponding to a spectral resolution of $R \sim 5{,}000$ (60~km~s$^{-1}$), the seeing was better than 1\arcsec, indicating an effective spectral resolution of $R \sim 10{,}000$ (30~km~s$^{-1}$). The H$\alpha$ line profile was reported to be an irregular shape, with a 10~\% width of $\sim$270~km~s$^{-1}$, corresponding to a line width of $\sim$150~km~s$^{-1}$ assuming a Gaussian profile. While variability was not examined in their study, a weak redshifted absorption component and a weak blueshifted emission peak may be present in the profile observed in 2016. If confirmed, this could indicate the presence of a variable yet long-lived redshifted absorption component and asymmetry. In addition, their inferred line width of $\sim$150~km~s$^{-1}$ is broader than our value of $\sim$100~km~s$^{-1}$. The H$\alpha$ line fluxes are comparable, assuming our flux calibration is reliable. Overall, long-term monitoring will be necessary to assess variability on yearly timescales.

Although the H$\alpha$ flux reported by \citet{Santamaria-Miranda2018SR12c} is comparable to that measured in this work, \citet{Santamaria-Miranda2018SR12c} detected multiple Balmer lines from H$\alpha$ to H12, whereas we detected only H$\alpha$. This difference is not necessarily surprising, as Balmer line ratios depend sensitively on the density, temperature, and velocity structure of the accretion flow \citep[e.g.,][]{Kwan2011Flow,Aoyama2018}. At higher gas densities, lower-order Balmer lines become increasingly optically thick, reducing the line ratios relative to higher-order transitions. The detection of multiple Balmer lines reported by \citet{Santamaria-Miranda2018SR12c} may therefore reflect a relatively dense accretion flow. In contrast, lower gas densities can lead to larger line ratios. In addition, because higher-order Balmer lines are generally more optically thin, they are expected to be more strongly affected by redshifted absorption \citep[see Figure~8 of][]{Muzerolle2001Model}, making them more difficult to detect. This may explain the non-detection of higher-order Balmer lines in our observations.

\section{Conclusions} \label{sec:conclusions}

We presented the first high-spectral-resolution spectroscopy of the wide-orbit planetary-mass companion SR~12~c at $R \sim 49,000$--40{,}000 (6.1--7.5~km~s$^{-1}$), aimed at investigating its accretion properties. Our observations consisted of five consecutive 30-minute integrations. H$\alpha$ emission was clearly detected, while higher-excitation Balmer lines (H$\beta$, H$\gamma$, and H$\delta$) were not detected, even after stacking all five datasets.

The H$\alpha$ line profiles in all five datasets were blueshifted and asymmetric with respect to \jh{the systemic velocity of SR~12~c (approximately $-7$~km~s$^{-1}$)}. We attributed the asymmetry to redshifted absorption or blueward asymmetry, which is a characteristic feature of magnetospheric accretion, a process widely regarded as the dominant accretion mechanism in young stars. Overall, the accretion process in SR~12~c appears to be star-like, consistent with magnetospheric accretion, \jh{while a scenario combining boundary-layer accretion with a failed wind cannot be ruled out.}

We also found variable line profiles. During the first hour, the line peak was located at $\sim -30$ km s$^{-1}$, whereas in subsequent observations the peak shifted to $\sim -10$ km s$^{-1}$. The line flux decreased over the first hour and then remained stable thereafter. Since our observations \jh{may not have captured the full range of variability}, the origin of the variability remained unclear. We discussed several possible scenarios, including rotational modulation, inhomogeneous accreting inflows, a binary-planet configuration, and a change in the dominant emission mechanism. Future observations, including high-angular-resolution imaging, time-resolved monitoring, and multi-line spectroscopy, will provide further insight into the accretion properties of SR~12~c.

\begin{acknowledgments}

The authors thank the anonymous referee for a timely and constructive report. 
This study was supported by JSPS KAKENHI Grant Number JP22KK0043, 23K03463, 25K07376, and JP26K00769.
This research is based [in part] on data collected at the Subaru Telescope, which is operated by the National Astronomical Observatory of Japan. We are honored and grateful for the opportunity of observing the Universe from Maunakea, which has the cultural, historical and natural significance in Hawaii. 
IRAF is distributed by the National Optical Astronomy Observatories, which are operated by the Association of Universities for Research in Astronomy, Inc., under cooperative agreement with the National Science Foundation.
M.T. is supported by the National Science and Technology Council (NSTC) of Taiwan (grant No. 114-2112-M-001-002-).
Y.A.\ acknowledges support from the National Science Foundation of China with grants No.~W2533003.
\end{acknowledgments}

\begin{contribution}

All authors contributed equally to the manuscript.


\end{contribution}

\software{
          IRAF \citep{Tody1986IRAF},
          gnuplot (\url{http://www.gnuplot.info}),
          }

\facilities{Subaru}

%

\bibliography{sample701}{}
\bibliographystyle{aasjournalv7}



\end{document}